# Solute softening and vacancy generation by diffusion-less climb in magnesium alloys

Peng Yi[1,2,*]

[1]Department of Materials Science and Engineering, [2]Hopkins Extreme Materials Institute, Johns Hopkins University, Baltimore, MD 21218, USA

*: corresponding author
Email: pengyi@jhu.edu

## Abstract

Active room temperature diffusion-less climb of the <a> edge dislocations in model Mg-Al alloys was observed using molecular dynamics simulations. Dislocations on prismatic and pyramidal I planes climb through the basal plane to overcome solute obstacles. This out-of-plane dislocation motion softens the high resistance pyramidal I glide and significantly reduces the anisotropy of dislocation mobility, and could help improve the ductility of Mg. The flow stress scales linearly with solute concentration, $c_{Al}$. Dislocations climb predominantly in the negative direction, with climb angle on the order of $0.01 c_{Al}$, producing very high vacancy concentration on the order of $10^{-4}$.

# 1 Introduction

Magnesium alloys have drawn increasing interests for their applications in automobile, aerospace, and defense industries as lightweight materials. However, their broad applications are hindered by the low ductility due to low HCP symmetry and strong anisotropy in deformation modes.[1] In an effort to improve their mechanical properties through additive manufacturing, computational studies on the atomistic level were devoted in recent years to study the dislocation and twinning mobilities, particularly the solute effects including strengthening, softening, dissociation, and multiplication.[2-12] Nevertheless, how to improve the capacity to better accommodate deformation in the *c*-axis still remain debated.(Figure 1(a))

Deformation in the *c*-axis requires participation of non-basal dislocations. In the past, much attention was paid to the <*c+a*> dislocations, and it was suggested that double cross-slip and multiplication of the <*c+a*> dislocations are the origin for improved ductility in certain alloys.[10, 12] However, the importance of non-basal <*a*> dislocation was still believed to play an important role,[13] and recently experimental work suggested that the climb of non-basal <*a*> dislocation can provide a satisfactory explanation for observed transitions in both anisotropy and texture evolution.[14]

Typically climb is closely related to creep, and is considered significantly active only at high temperature, facilitated by the diffusion of vacancies.[15-17] For example, the temperature range for creep in Mg is 600-780K [18-20]. The room temperature climb, if occurs, is not likely due to the diffusion of vacancies. Instead, it could potentially be driven by deformation and serves as source of vacancies. The ability to generate large number of vacancies makes climb not only a mechanically interesting feature, e.g. voids and fracture, but also an important subject from materials processing point of view, where excess of vacancies stimulate diffusion, precipitation, segregation, or recrystallization.[21]

# 2 Methods

We study the glide of the <*a*> dislocations on two non-basal planes, prismatic and pyramidal I planes, and demonstrated that the diffusion-less climb can be induced by solute atoms.

Molecular static (MS) and Molecular dynamics (MD) simulations were performed using LAMMPS.[22] Mg alloys were modeled using Kim *et al.*'s MEAM potentials for Mg/Al [23]. This MEAM potential has proven to yield good agreement with DFT calculations of more complicated <c+a> dislocation core structures in Mg[24], and has been used for simulating the solute effects on dislocation glide and twin propagation in our previous studies.[5, 7, 11] MS simulations were performed by energy minimization using the conjugate gradient method with energy tolerance being $10^{-14}$ throughout. For MD simulations, the system was kept thermally isolated, i.e., no thermostat/barostat was used to minimize the addition of spurious dissipative dynamics. Since the maximum shear strain applied to the system in our simulation did not exceed 0.05, the temperature rises due to the shear work was negligible. Time was integrated using Verlet algorithm with a time step of 2 fs. The Common Neighbor Analysis method [25, 26] with a cutoff of 3.8 Å was used to identify the atoms in defects.

The simulation box contains one positive edge dislocation. The Burgers vector is along *x*-direction, and the dislocation line is along *y*- direction. The lattice constants are obtained through separate equilibration simulations at given temperature and solute concentration at 0 Pa. The box dimension is about 32 nm in *x*- and *z*-directions and 16 nm in *y*-direction, thus the dislocation density is $10^{15} m^{-2}$, which is comparable to that in severely deformed metals.[27] The system volume remains constant, and there are about $7\times 10^5$ atoms in the system. The Periodic boundary conditions were applied in *x*- and *y*-directions, and a mixed boundary condition was applied to the *z*-direction, similar to a previous work.[5, 7, 11] Specifically, two slabs were created at the ±*z* boundaries with thickness $d_{slab}$, chosen to be greater than the cutoff distance of the potential. Atoms in each slab comprise a rigid body. In addition, both slabs were prohibited from moving in the *z*-direction throughout the simulations. Simple shear with constant strain rates $5\times 10^7$ s$^{-1}$ was simulated by displacing the +*z* slab in *x*-direction, while the –*z* slab was held fixed. To study the solute effect, atoms between the two slabs were randomly substituted with solute atoms. Temperatures from 0K to 500K were studied, as were solute concentration ranges from Al 0-7 at.%. Open source software VMD [28] was used for visualization.

## 3    Results and discussion

The <a> edge dislocation is stable on the prismatic plane, unlike the <a> screw dislocation which tends to cross-slip to the basal plane.[7] The representative stress-strain curves of an <a> edge dislocation gliding on the prismatic plane at 300K are shown in Figure 1(b) for different Al concentrations, $c_{Al}$. The flow stress, or the CRSS, is calculated as the average value after the stress-strain curve reaches plateau after yield.

The CRSS of Mg alloys is plotted as a function of solute concentration in Figure 1(c). Excluding the data for pure Mg, the CRSS scales linearly with $c_{Al}$ for all temperatures in this study, suggesting a consistent mechanism for dislocation motion in the alloys. This scaling is consistent with our previously study of basal <a> dislocation glide, as well as experiments, where the scaling transitions from $c_{Al}^{2/3}$ at low temperatures to $c_{Al}^1$ at high temperatures.[5, 29] The $c_{Al}^{2/3}$ scaling is predicted by Labusch statistics [30], and the linear scaling is predicted by the Suzuki-Ishii development [31] of the Fleischer-Friedel model. However, the obstacle strength and the dislocation line energy could significantly affect the scaling [30], as also observed in Ni/Al and Al/Mg alloy systems.[32, 33]

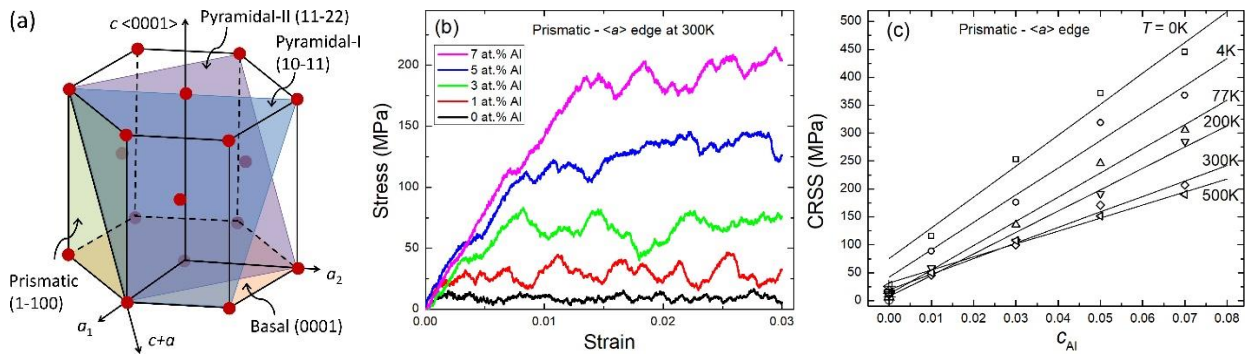

Figure 1  (a) Crystal structure and slip planes in Mg. (b) Representative stress-strain curves for simple shear applied to a crystal containing an <a> edge dislocation on prismatic plane at 300K for different solute concentrations. (c) Solute concentration dependence of CRSS of prismatic glide of <a> edge dislocation for different temperatures. Symbols are simulation data, where center-crossed ones are for pure Mg. Lines are linear fitting results using data for Mg alloys (excluding pure Mg).

Diffusion-less climb was observed in Mg alloys. Snapshots of a gliding prismatic <a> edge dislocation in an alloy with 1 at.% Al at 300K is shown in Figure 2(a)-(d), where Mg and Al atoms are shown in red and blue, respectively. A short energy minimization was performed to reduce thermal noises, and only defect atoms are shown. At time $t_1$ (Figure 2(a)(b)), several solute atoms serve as a barrier to pin the dislocation. With increasing shear stress, the barrier was overcome in the next 50 ps, by dislocation climb. Figure 2(c) shows the unpinned dislocation and two small clusters of atoms. The atoms in the clusters all have the number of nearest neighbors equal to or less than 11, indicating that that they are surrounding one or more vacancies. Figure 2(d) shows that the jog pair distance is 2$c$, where $c$ is the unit cell dimension along the <0001> direction (Figure 1(a)). The jog has a height of about 0.27 nm, which corresponds to the distance between two neighboring prismatic planes. Since this segment climbs in the negative direction, 4 vacancies were produced, as shown in Figure 2(c). The 4 vacancies are not produced all at once, as the dislocation core is able to retain single site defects.

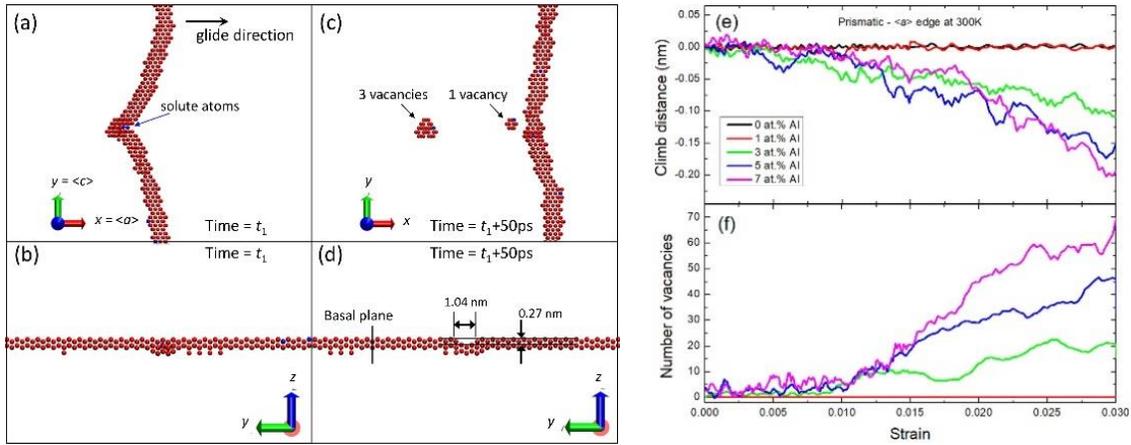

Figure 2 Snapshots of an <a> edge dislocation gliding on the prismatic plane through solute induced climb in Mg alloy of 1 at.% Al at 200K. (a)(b) Dislocation is pinned by solute atoms at time $t_1$, (a) views into the glide plane and (b) views into the glide direction. (c)(d) Dislocation is unpinned through negative climb, generating 4 vacancies behind. (d) only shows dislocation for clarity. (e)(f) Climb distance in the $z$-direction and the number of vacancies as functions of strain at 300K (5-point average was used to reduce noises).

No jog was found at any temperature for pure Mg, suggesting that the thermal jog density in pure Mg is low. Climb, i.e., out-of-plane motion, requires high thermal activation energy and must be induced by solute atoms. In addition, no jog was found at 4K and 77K for Mg-Al 1 at.% alloy, suggesting that the in-plane Pierels barrier is lower than the formation energy of the jog pair under these conditions. Typically climb is active only at high temperature, induced by diffusion of pre-existing vacancies in the crystal. Diffusion-less climb process was observed for <c+a> dislocation glide in pure Mg using Liu's EAM potential [34] in compression simulations at very low temperature [35]. However, the climb observed there was due to free surfaces. Our study shows that solute induced climb of <a> dislocation in the crystal bulk at room temperature is very active.

The <a> dislocation climbs in the negative direction because the formation energy of a vacancy is lower than that of an interstitial, and act as vacancy sources. The average z-coordinate of the dislocation and the number of vacancies are plotted as functions of strain for different Al concentrations at 300K in Figure 2(e) and (f). The climb angle is defined as $dz/dx$, where $dz$ is the climb distance along the z-direction, and $dx$ is the dislocation glide distance and is calculated using the Orowan's equation, $\varepsilon=\rho b dx$. The linear fitting of Figure 2(e) gives the climb angle to be $dz/dx \cong -0.03 c_{Al}$. On the other hand, the vacancy concentration after stain $\varepsilon=0.03$ is on the order of $10^{-4}$. This very high vacancy concentration is also comparable to that observed in experiments for severely deformed metals.[27, 36] These values are far exceeding the typical equilibrium vacancy concentration based on formation energy.[37] Therefore, current vacancy generation model based on thermal jogs need to be revised to account for the much enhanced jog creation rates especially in alloys.[21]

On pyramidal I plane, solute softening was observed for <a> edge dislocation glide. Figure 3(a) shows the representative stress-strain curves at 300K, where the flow stresses (CRSS) for 1 and 3 at.% Al are significantly lower than that for pure Mg. The solute softening observed in Figure 3(a) is also due to climb, as will be discussed shortly. The CRSS of pyramidal I glide as a function of temperature is shown in Figure 3(b), for different solute concentrations. Solute softening was observed for temperatures up to 500K. The temperature dependence of the flow stress is similar to what was observed experimentally in Fe alloys [38]. Both cases have high lattice resistance against in-plane glide, and out-of-plane dislocation motion (cross-slip in Fe and climb in Mg) is activated to help reduce the flow stress. In Mg, this softening significantly reduces the anisotropy in dislocation mobility. For example, the ratio of CRSS for <a> edge dislocation on basal, prismatic and pyramidal I planes at 300K changes from about 1:2:19 for pure Mg to 1:3:4 for alloys with 3 at.% Al.[5]

The CRSS for glide on the pyramidal I plane as a function of solute concentration for different temperatures is plotted in Figure 3(c). Excluding the data for pure Mg, the CRSS scales linearly with $c_{Al}$ consistently, suggesting a consistent mechanism for dislocation motion in the alloys, same as the prismatic glide.

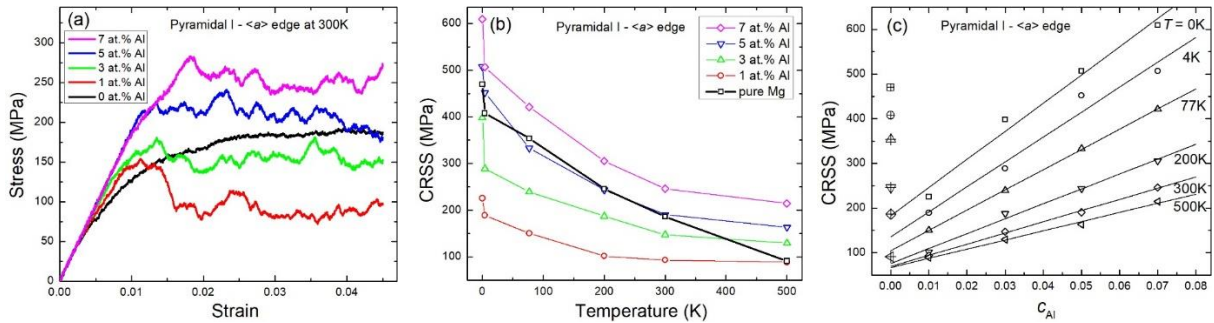

Figure 3. (a) Representative stress-strain curves for simple shear applied to a crystal containing an <a> edge dislocation on pyramidal I plane at 300K for different solute concentrations. (b) CRSS as a function of temperature for different solute concentrations (Lines are guide for the eye). (c) CRSS as a function of solute concentration for different temperatures. Symbols are simulation data, where center-crossed ones are for pure Mg. Lines are linear fitting results using data for Mg alloys (excluding pure Mg).

Our simulation shows that dislocation moves in-plane in *pure* Mg, but it overcomes solute obstacles by climb in alloys. Figure 4(a) and (b) show a snapshot of an <a> edge dislocation gliding on pyramidal I plane in an Mg alloy with 1 at.% Al at 300K, where Mg and Al atoms are shown in red and blue, respectively. Figure 4(a) shows atoms in dislocation and the atoms surrounding defects. These defects are all vacancies, as confirmed by the number of nearest-neighbor of these atoms. The vacancies to the right of the dislocation were generated by itself in the periodic image box. Figure 4(b) shows only the dislocation with three jog pairs. The height of the jogs is about 0.23nm, corresponding to the distance between nearest pyramidal I plane pairs, which each pair consists of two planes 0.04nm apart.[35] There are 4 basal planes between each jog pair. Climb of the pyramidal I <a> edge dislocation creates an <a> mixed dislocation segment on the basal plane, which has much lower flow stress than the pyramidal I dislocations.

This more mobile basal segment contributes to the reduced flow stress and the observed solute softening.

Same as the prismatic glide, the <a> dislocation climbs in the negative direction, acting as vacancy sources. The average z-coordinate of the dislocation and the number of vacancies are plotted as functions of strain for different Al concentrations at 300K in Figure 4(c) and (d). The plateaus for alloys with 1 and 3 at.% Al are due to the dislocation gliding into the periodic image of the vacancy field itself generates, as explained above. The climb angle $dz/dx \cong -4c_{Al}$, which is about twice as much as the prismatic slip. This is probably due to the higher tendency to climb for pyramidal slip because of high shear stress. As a result, the vacancy generation rate is also about twice as much as the prismatic slip.

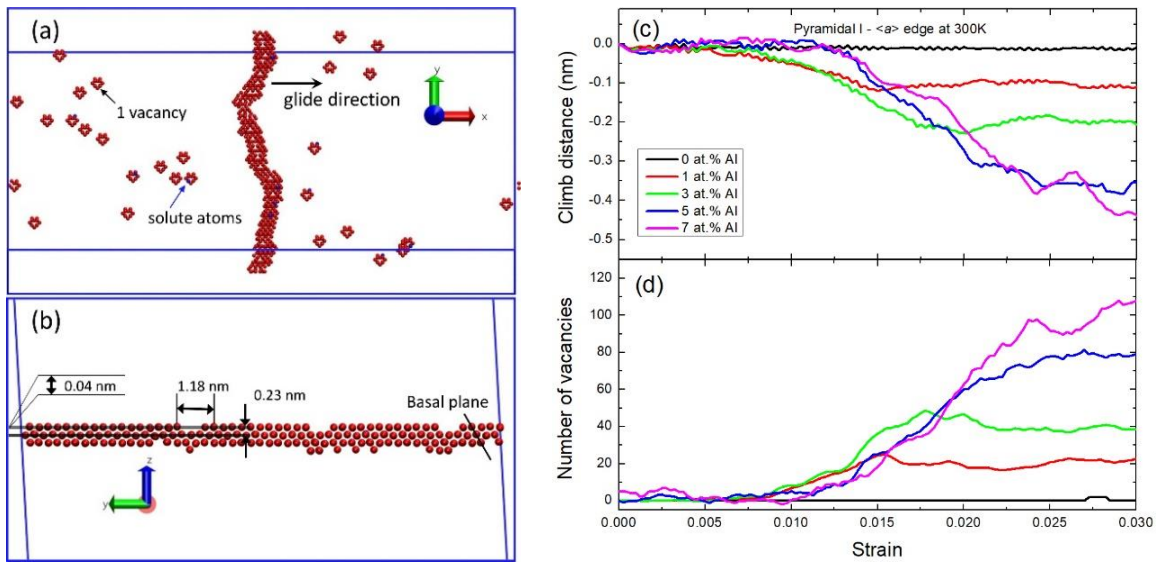

Figure 4. (a) Snapshot of an <a> edge dislocation gliding on pyramidal-I in Mg alloy of 1 at.% Al at 300K (Red is for Mg atoms and blue for Al atoms.): (a) View into the glide plane, and (b) view into the glide direction (only dislocation is shown for clarify). See text for details. (c)(d) Climb distance in the z-direction and the number of vacancies as functions of strain at 300K (5-point average was used to reduce noises).

## 4  Conclusion

In conclusion, we have observed and investigated the active climb of $<a>$ edge dislocation when gliding on prismatic and pyramidal I planes in Mg alloys.  Climb cause softening pyramidal I glide and helps reduce the anisotropy in CRSS, which could potentially help improve the ductility.  In addition, the climb is dominantly along the negative direction, producing a very high vacancy concentration, $10^{-4}$.  These excess vacancies could facilitate the climb of other dislocations, for example, $<c>$ dislocations, to further improve the ductility of the Mg alloys.  On the other hand, they could also play important role in other plasticity and failure problems, as well as phase transformation problem like dynamic materials processing.  It is worth noting that we applied a high strain rate of $5\times10^7$ s$^{-1}$ in this study.  It is expected that with lower strain rate, the climb will be more active as the equivalent temperature is even higher.  This solute induced climb mechanism needs to be accounted for in multiscale modeling of plasticity and phase transformation.  For this purpose, accurate first principle calculation on the energetics of solute induced climb is underway.

*Acknowledgments*

This work was sponsored by the Army Research Laboratory and was accomplished under Cooperative Agreement Number W911NF-12-2-0022.  The views and conclusions contained in this document are those of the authors and should not be interpreted as representing the official policies, either expressed or implied, of the Army Research Laboratory or the U.S. Government.  The U.S. Government is authorized to reproduce and distribute reprints for Government purposes notwithstanding any copyright notation herein.  Computational resources from Maryland Advanced Research Computing Center (MARCC) are acknowledged.

*References*:

1. Zhang, J. and S.P. Joshi, *Phenomenological crystal plasticity modeling and detailed micromechanical investigations of pure magnesium.* Journal of the Mechanics and Physics of Solids, 2012. **60**(5): p. 945-972.
2. Yasi, J.A., L.G. Hector Jr, and D.R. Trinkle, *Prediction of thermal cross-slip stress in magnesium alloys from a geometric interaction model.* Acta Materialia, 2012. **60**(5): p. 2350-2358.
3. Tang, Y. and J.A. El-Awady, *Highly anisotropic slip-behavior of pyramidal I 〈c+a〉 dislocations in hexagonal close-packed magnesium.* Materials Science and Engineering: A, 2014. **618**(0): p. 424-432.
4. Itakura, M., et al., *Atomistic study on the cross-slip process of a screw <a> dislocation in magnesium.* Modelling and Simulation in Materials Science and Engineering, 2015. **23**(6): p. 065002.
5. Yi, P., R.C. Cammarata, and M.L. Falk, *Atomistic simulation of solid solution hardening in Mg/Al alloys: Examination of composition scaling and thermo-mechanical relationships.* Acta Materialia, 2016. **105**: p. 378-389.
6. Wu, Z. and W.A. Curtin, *Intrinsic structural transitions of the pyramidal I 〈c + a〉 dislocation in magnesium.* Scripta Materialia, 2016. **116**: p. 104-107.
7. Yi, P., R.C. Cammarata, and M.L. Falk, *Solute softening and defect generation during prismatic slip in magnesium alloys.* Modelling and Simulation in Materials Science and Engineering, 2017. **25**(8): p. 085001.
8. Fan, H., et al., *Temperature effects on the mobility of pyramidal < c + a > dislocations in magnesium.* Scripta Materialia, 2017. **127**: p. 68-71.
9. Tehranchi, A., B. Yin, and W.A. Curtin, *Solute strengthening of basal slip in Mg alloys.* Acta Materialia, 2018. **151**: p. 56-66.
10. Wu, Z., et al., *Mechanistic origin and prediction of enhanced ductility in magnesium alloys.* Science, 2018. **359**(6374): p. 447-452.
11. Yi, P. and M.L. Falk, *Thermally activated twin thickening and solute softening in magnesium alloys - a molecular simulation study.* Scripta Materialia, 2019. **162**: p. 195-199.
12. Ahmad, R., et al., *Designing high ductility in magnesium alloys.* Acta Materialia, 2019. **172**: p. 161-184.
13. Agnew, S.R., *2 - Deformation mechanisms of magnesium alloys*, in *Advances in Wrought Magnesium Alloys*, C. Bettles and M. Barnett, Editors. 2012, Woodhead Publishing. p. 63-104.
14. Ritzo, M.A., et al., *An Investigation into the Role of Dislocation Climb During Intermediate Temperature Flow of Mg Alloys*, in *Magnesium Technology 2020*. 2020. p. 115-122.
15. Caillard, D., *Thermally activated mechanisms in crystal plasticity*, ed. J.-L. Martin. 2003, Amsterdam: Pergamon.
16. Kabir, M., et al., *Predicting dislocation climb and creep from explicit atomistic details.* Phys Rev Lett, 2010. **105**(9): p. 095501.
17. Clouet, E., *Predicting dislocation climb: Classical modeling versus atomistic simulations.* Physical Review B, 2011. **84**(9).
18. Edelin, G. and J.P. Poirier, *Etude de la montée des dislocations au moyen d'expériences de flu age par diffusion dans le magnésium.* Philosophical Magazine, 1973. **28**(6): p. 1203-1210.
19. Edelin, G. and J.P. Poirier, *Etude de la montée des dislocations au moyen d'expériences de fluage par diffusion dans le magnésium.* Philosophical Magazine, 1973. **28**(6): p. 1211-1223.
20. Vagarali, S.S. and T.G. Langdon, *Deformation mechanisms in h.c.p. metals at elevated temperatures—I. Creep behavior of magnesium.* Acta Metallurgica, 1981. **29**(12): p. 1969-1982.

21. Militzer, M., W.P. Sun, and J.J. Jonas, *Modelling the effect of deformation-induced vacancies on segregation and precipitation.* Acta Metallurgica et Materialia, 1994. **42**(1): p. 133-141.
22. Plimpton, S., *Fast Parallel Algorithms for Short-Range Molecular Dynamics.* Journal of Computational Physics, 1995. **117**(1): p. 1-19.
23. Kim, Y.-M., N.J. Kim, and B.-J. Lee, *Atomistic Modeling of pure Mg and Mg–Al systems.* Calphad, 2009. **33**(4): p. 650-657.
24. Ghazisaeidi, M., L.G. Hector Jr, and W.A. Curtin, *First-principles core structures of edge and screw dislocations in Mg.* Scripta Materialia, 2014. **75**(0): p. 42-45.
25. Faken, D. and H. Jónsson, *Systematic analysis of local atomic structure combined with 3D computer graphics.* Computational Materials Science, 1994. **2**(2): p. 279-286.
26. Tsuzuki, H., P.S. Branicio, and J.P. Rino, *Structural characterization of deformed crystals by analysis of common atomic neighborhood.* Computer Physics Communications, 2007. **177**(6): p. 518-523.
27. Zehetbauer, M., et al., *Deformation Induced Vacancies with Severe Plastic Deformation: Measurements and Modelling.* Materials Science Forum, 2006. **503-504**: p. 57-64.
28. Humphrey, W., A. Dalke, and K. Schulten, *VMD: Visual molecular dynamics.* Journal of Molecular Graphics, 1996. **14**(1): p. 33-38.
29. Akhtar, A. and E. Teghtsoonian, *Substitutional solution hardening of magnesium single crystals.* Philosophical Magazine, 1972. **25**(4): p. 897-916.
30. Leyson, G.P.M. and W.A. Curtin, *Friedel vs. Labusch: the strong/weak pinning transition in solute strengthened metals.* Philosophical Magazine, 2013. **93**(19): p. 2428-2444.
31. Suzuki, T. and T. Ishii, *Physics of strength and plasticity*, ed. A.S. Argon. 1969, Cambridge: M.I.T. Press.
32. Patinet, S. and L. Proville, *Depinning transition for a screw dislocation in a model solid solution.* Physical Review B, 2008. **78**(10): p. 104109.
33. Patinet, S. and L. Proville, *Dislocation pinning by substitutional impurities in an atomic-scale model for Al(Mg) solid solutions.* Philosophical Magazine, 2011. **91**(11): p. 1581-1606.
34. Liu, X.-Y., et al., *EAM potential for magnesium from quantum mechanical forces.* Modelling and Simulation in Materials Science and Engineering, 1996. **4**(3): p. 293.
35. Tang, Y. and J.A. El-Awady, *Formation and slip of pyramidal dislocations in hexagonal close-packed magnesium single crystals.* Acta Materialia, 2014. **71**(0): p. 319-332.
36. Čížek, J., et al., *The Development of Vacancies during Severe Plastic Deformation.* MATERIALS TRANSACTIONS, 2019. **60**(8): p. 1533-1542.
37. Hull, D., *Introduction to Dislocations*, D.J. Bacon, Editor. 2011, Elsevier Science: Burlington :.
38. Takeuchi, S., H. Yoshida, and T. Taoka, Trans. Japan Inst. Metals, (Supplement), 1968. **9**: p. 715.